\documentclass[12pt]{article}
\usepackage{amsmath,amssymb,amsthm,color}
\usepackage{graphicx}
\usepackage{cite}
\usepackage{epsfig}
\renewcommand{\baselinestretch}{2}
\begin{document}
%
\title{Beating oscillations of magneto-optical spectra in simple hexagonal graphite \\}
\author{
\small Rong-Bin Chen$^{a}$, Yu-Huang Chiu$^{b,c,*}$, Ming-Fa Lin$^{b,*}$ $$\\
\small  $^a$Center of General Studies, National Kaohsiung Marine
University, Kaohsiung 811, Taiwan\\
\small  $^b$Department of Physics, National Cheng Kung University, Tainan 701, Taiwan \\
\small  $^c$Department of Applied Physics, National Pingtung University, Pingtung 900, Taiwan \\
 }
\renewcommand{\baselinestretch}{1}
\maketitle

\renewcommand{\baselinestretch}{1.4}
\begin{abstract}

The magneto-optical properties of simple hexagonal graphite exhibit rich
beating oscillations, which are dominated by the field strength and photon
energy. The former has a strong effect on the intensity, the energy range of
the beating and the number of groups, and the latter modulates the total group numbers of the oscillation structures. The single-particle and collective excitations
are simultaneously presented in the magnetoreflectance spectra and can
be precisely distinguished. For the loss function and reflectance, the
beating pattern of the first group displays stronger intensities and broader
energy range than other groups. Simple hexagonal graphite possesses unique
magneto-optical characteristics that can serve to identify other bulk graphites.
\vskip 1.0 truecm
\par\noindent

\vskip1.0 truecm

\par\noindent  * Corresponding author. {~ Tel:~ +886-6-2159492.}\\~{{\it E-mail addresses}: airegg.py90g@nctu.edu.tw (Y.H. Chiu), mflin@mail.ncku.edu.tw (M.F. Lin)}
\end{abstract}

\pagebreak
\renewcommand{\baselinestretch}{2}
\newpage

{\bf 1. Introduction}
\vskip 0.3 truecm

Graphene-based materials have attracted many researchers to investigate
their physical properties due to their potential for novel applications. The
material properties demonstrate their strong dependence on the stacking
configurations \cite{KFMak, AGruneis, JAdler, SYuan, MKoshino,YLiu}, layer numbers \cite{BPartoens, TKaplas, YHao, SCheon, OLBerman, ZSWu}, and interlayer atomic interactions \cite{JCCharlier, LSamuelson, JCCharlier2}. Graphite is composed of van der Waals coupled graphene layers \cite{DDLChung, KBerland}. Three prototypes of periodical stacking along the $z $-direction exist: AA-stacked simple hexagonal graphite (SHG), AB-stacked Bernal
graphite (BG), and ABC-stacked rhombohedral graphite (RHG). The graphites
discovered in nature are mainly composed of BG and RHG. Recently, SHG has
been successfully synthesized in the laboratory \cite{JKLee}. The
interlayer couplings play an important role in determining the low-energy
electronic properties; thus, different periodic stacking configurations
exhibit their own unique characteristics. It is known that the neighboring
electronic states congregate and form the Landau subbands (LSs)
along $\widehat{k_{z}}$ in a perpendicular uniform magnetic field \textbf{B%
}=$B_{0}\widehat{z}$. The magneto-electronic properties demonstrate very
interesting phenomena, e.g., the anisotropy of \ low-energy electronic
structures \cite{DDLChung, RBChen, YHHo, CHHo}, the de Haas-van Alphen effect \cite{JWMcClure,IALukyanchuk}, quantum Hall effect \cite{IALukyanchuk2, BABernevig, ANRamanayaka, YZheng}. In this work, we
mainly focus on obtaining the magneto-optical properties of SHG by means of
evaluating the dielectric function $\varepsilon (\omega ,B_{0})$.
Comparisons with BG and RHG are also made.

The LSs of graphites present many important features. SHG possesses
very strong $k_{z}$-dependent energy dispersions with a broad band width about 1
eV, and each LS can be described by a simple relationship with $k_{z}$ \cite{RBChen, BABernevig}.
Many LSs cross the Fermi level ($E_{F}=0$) \cite{RBChen, RBChen2}.
Moreover, the excitations related to the densely low-lying LSs own wide energy ranges which can overlap for different optical transition channels. On the contrary, RHG
exhibits weak $k_{z}$-dependent dispersions with a narrow band width ($\sim $%
10 meV) \cite{CHHo}. Only one LS crosses $E_{F}$, and there is no coexisting energies for different optical excitations. The LS
can be characterized by the approximate solution \cite{CHHo2}. The energy
dispersion of BG has a band width of $\sim $0.2 eV, which lies between that of
SHG and RHG, and two LSs cross $E_{F}$ \cite{KCChuang, YHHo2}. The low-lying LSs are complex
and cannot be easily described by $k_{z}$. The characteristics of LSs would
be reflected in the magneto-optical spectra.

The magneto-optical properties are closely associated with the stacking
configurations of graphites \cite{YHHo2, RBChen2, TMatsui, JMSchneider}. The low-energy magneto-optical absorption spectrum of SHG is dominated by intraband and interband optical excitations which induce
a multi-channel threshold peak, several two-channel peaks, and many
double-peak structures \cite{RBChen2}. In the magneto-optical absorption
spectra of BG, the prominent peaks originate from the interband excitations
at both the $K$ and $H$ points. The peaks associated with the $K$ point\ display
double-peak structures \cite{KCChuang, YHHo}. Moreover,
the field evolution of the absorption lines for the $K$-point type shows an approximately
 linear dependence, while the dependence of the $H$ point is square-root like \cite{MOrlita, YHHo, KCChuang}. The magnetoreflectance $R(\omega ,B_{0})$ spectra of BG displays irregular oscillations \cite{ZQLi, LCTung}. Up to now,
no theoretical calculations and experimental measurements for the magneto-optical absorption or reflection of {\small RHG} have been performed {\small . }

The magneto-optical properties are evaluated based on the Peierls
tight-binding model, which can be exactly diagonalized even with the
inclusions of field-induced Peierls phases and important atomic interactions in
the Hamiltonian \cite{YHHo2, RBChen2, YHLai}. This study shows
that the beating patterns of the dielectric function can be formed
mainly owing to the strong overlap of different optical transition
channels in a wide frequency range. Such beating patterns are also exhibited in the
higher-frequency absorption spectrum, loss function, and reflectance. The
single-particle and collective excitations can be precisely identified,
respectively, based on the shoulders (peaks) and dips of specific structures in the
magnetoreflectance spectra. The regular beating
magneto-optical spectra can be controlled by the field strength and the
photon energy, which provide a theoretical basis for future experiments to
clarify the optical responses of the graphite configurations.

The generalized tight-binding model deserves a closer examination in numerical calculations. We developed this model to study the magneto-electronic and -optical properties by the exact diagonalization method. In studying the magneto-electronic properties, the earlier work can only cope with eigenvalues and eigenfunctions at strong magnetic field strength \cite{CPChang} because the Hamiltonian matrix gets too large as the field strength decreases. For example, this matrix is 40000 $\times$ 40000 for monolayer graphene at 7.9 T. By means of rearranging the tight-binding functions, it is possible to transform the huge matrix into a band-like one. Therefore, the eigenvalues and the wave functions can be efficiently solved at weaker field strength ($\sim $ 1T) \cite{YHLai}. In this work, the magneto-optical absorption spectra, which are determined by three large matrices due to the Hamiltonian, the initial state and the final state, can be obtained by using the localized features of the magnetic wave functions. The PC clusters are sufficient in calculating numerical data. The acquired features of LS spectra and the reliable characterization of the LS wave functions provide a guideline for other physical properties, such as Coulomb excitations and transport properties. As for the discussion of the optical properties in our previously published works, the generalized tight-binding method has been successfully applied to investigate the magneto-optical absorption spectra of few-layer graphenes. The optical selection rules are well defined through the detailed analysis on the wave functions. It is also applicable to bulk graphite with layers stacked in any sequence. Furthermore, the generalized tight-binding model can be used in the cases of spatially modulated fields and combined magnetic and electric fields.
\vskip 0.6 truecm
\par\noindent
{\bf 2. Methods }
\vskip 0.3 truecm

For calculation purposes, the geometric structure of simple hexagonal graphite is regarded as a
stacking sequence of infinite layers of graphene with an AA-stacked
configuration along $\widehat{z}$. All honeycomb structures in SHG have the
same projections on the x-y plane.
The interlayer distance is $I_{c}
=3.50$ \AA \thinspace\ \cite{JKLee} and the C-C bond length is $b^{\prime }=1.42$
\AA. A primitive unit cell consists of two atoms. The associated hopping
integrals $\gamma _{i}$'s taken into account are the one
intralayer atomic interaction ($\gamma _{0}$ $=$ $2.519$ eV) and three
interlayer atomic interactions ($\gamma _{1}$ $=$ $0.361$ eV; $\gamma _{2}$ $%
=$ $0.013$ eV; $\gamma _{3}$ $=$ $-0.032$ eV) \cite{JCCharlier2}.

 When SHG is subjected to a $B_{0}\widehat{z}$, the path integral of the
vector potential induces a periodical Peierls phase (details in Ref. [19]).
The phase term of the associated period is inversely proportional to the
magnetic flux ($\Phi =3\sqrt{3}b^{\prime 2}B_{0}/2$) through a
hexagon. To satisfy the integrity of the primitive cell, the ratio $R_{B}=\Phi _{0}$/$\Phi $\ ($\Phi _{0}$\ $%
=$\ $hc/e$ flux quantum) has to be a positive integer. As a result, the extended rectangular unit cell has $%
4R_{B}$ carbon atoms. The $\pi $-electronic Hamiltonian built from the $4R_{B}$
tight-binding functions is a $4R_{B}\times 4R_{B}$ Hermitian matrix. To
solve this huge matrix problem, one can convert the Hamiltonian matrix into a
band-like form by rearranging the tight-binding functions \cite{RBChen, CHHo2, YHHo2}.
Both eigenvalue $E^{c,v}$and eigenfunction $\Psi ^{c,v}$are efficiently
obtained, even for a small magnetic field. The superscripts $c$\ and $v$,
respectively, represent the conduction and valence bands.

The main features of the electronic properties can be directly manifested by
optical excitations. As materials absorb photons, electrons are excited from
occupied states to unoccupied states. Within the relaxation-time
approximation \cite{LGJohnson}, the transverse dielectric function at zero temperature is
expressed as \

\begin{eqnarray}
\varepsilon (\omega ,B_{0}) &=&\epsilon _{0}-\frac{e^{2}}{\pi ^{2}}\underset{%
n,n^{\prime }}{\sum }\underset{h,h^{\prime }=c,v}{\sum }\int_{1st\text{ }%
BZ}d^{3}\mathbf{k}\frac{\left\vert \left\langle \Psi _{n^{\prime
}}^{h^{^{\prime }}}(\mathbf{k})\left\vert \frac{\widehat{E}\cdot \mathbf{P}}{%
m_{e}}\right\vert \Psi _{n}^{h}(\mathbf{k})\right\rangle \right\vert ^{2}}{%
\omega _{hh^{\prime }}^{2}(n,n^{\prime };\mathbf{k})}  \notag \\
&&\times {\{}\frac{1}{\omega -\omega _{hh^{\prime }}(n,n^{\prime };\mathbf{%
k)+}i\Gamma }-\frac{1}{\omega +\omega _{hh^{\prime }}(n,n^{\prime };\mathbf{%
k)+}i\Gamma }\},
\end{eqnarray}%
where $\epsilon _{0}=2.4$ is the background dielectric constant \cite{EATaft}. $\omega
_{hh^{\prime }}(n,n^{\prime };\mathbf{k})=E^{h^{\prime }}(n^{\prime },%
\mathbf{k})-E^{h}(n,\mathbf{k})$ is the optical excitation energy which comes from the intraband $%
(c\rightarrow c;v\rightarrow v)$ or interband excitations $(v\rightarrow c)$%
; $\Gamma (=3.5$ meV$)$ is the broadening parameter due to the deexcitation
mechanisms. In these optical excitations, the momentum of the photons is
nearly zero and thus the excitations can be regarded as a vertical
transition between two LSs. The initial and final states have the same
wavevector, i.e., $\triangle k_{x}=0$, $\triangle k_{y}=0$, and $\triangle
k_{z}=0$ \cite{RBChen2}. The velocity matrix element $D_{m}=\left\langle \Psi
_{n^{\prime }}^{h^{^{\prime }}}(\mathbf{k})\left\vert \frac{\widehat{E}\cdot
\mathbf{P}}{m_{e}}\right\vert \Psi _{n}^{h}(\mathbf{k})\right\rangle $ is
evaluated within the gradient approximation \cite{LGJohnson, CWChiu}. As $\left\vert
D_{m}\right\vert ^{2}/\omega _{hh^{\prime }}^{2}$ is set to be a constant,
the imaginary part of $\varepsilon (\omega ,B_{0})$ is simply the joint
density of states D$_{J}(\omega ,B_{0})$. The evaluation of $\varepsilon
(\omega ,B_{0})$ can be employed to study the absorption spectrum, loss
function, and reflectance.

\vskip 0.6 truecm
\par\noindent
{\bf 3. Results and discussion}
\vskip 0.3 truecm

 The perpendicular magnetic field causes the cyclotron motion in the ${x}$-${y}$
plane; therefore, the Landau levels lie on the $k_{x}$-$k$$_{y}$
plane and the LSs along $\widehat{k}_{z}$. The energy dispersions of the LSs
along the $K-H$ line $(0\leq k_{z}(\pi /I_{\text{c}})\leq 1)$ exhibit a strong
dependence on $k_{z}$, as shown in Fig. 1. Based on the node structure of
the Landau wave functions, the quantum number $n^{c}(n^{v})$ for each
conduction (valence) LS can be identified by the total number of nodes \cite{RBChen2}.
 The LSs with $n^{c}$ and those with $n^{v}$ are asymmetric about ${E_{F}=0}$
  because of the interlayer atomic interactions. In
optical excitations, electrons are excited from occupied LSs into
unoccupied LSs. For the sake of convenience, the excitations between two
LSs with quantum numbers ${n^{c,v}}$ and ${m^{c,v}}$ are represented as $[n^{c(v)},m^{c(v)}]$ and $(n^{v},m^{c})$ for intraband and interband excitations, respectively. Moreover, the wave functions of occupied and unoccupied states
offer important insights into the possible excitation channels.
Since the LS wave functions of SHG are similar to those of monolayer graphene,
the same selection rule $\left\vert \triangle n\right\vert
=\left\vert m^{c,v}-n^{c,v}\right\vert =1$ applies \cite{RBChen2, MKoshino2, MLSadowski}.

 To investigate the spectrum structure of SHG, an illustration of
optical excitations is exhibited in Fig. 1 to show the existence of intraband and interband
optical excitations. The two intraband excitations $[n^{c},(n+1)^{c}]$ and $%
[(n+1)$$^{v},n^{v}]$ exhibit a tiny frequency discrepancy, and similar
results are also obtained for the two interband excitations $($\ $n^{v},(n+1)^{c})$
and $((n+1)$$^{v},n^{c})$. The former two and latter two can be simplified as $%
[n^{{}},n+1]$ and $(n^{{}},n+1)$, respectively. The intersection point of
each LS with the Fermi level is the Fermi-momentum state, $k_{F}^{n^{c,v}}$.
The effective $k_{z}$-range is confined by the initial and the final
Fermi-momentum states, i.e., \ $k_{F}^{n^{c,v}}\leq $ $k_{z}\leq
k_{F}^{(n\pm 1)^{c,v}}$, as shown in Fig. 1 by the colored arrows. The
intraband excitations with smaller $n^{c(v)}$'s possess a broader effective $%
k_{z}$-range (Fig. 1(a)). This reflects the fact that a quick
decline of the energy spacing between two adjacent LSs due to increasing
the quantum number $n^{c(v)}$'s. The effective $k_{z}$-ranges related to the
interband excitations are larger than those of the intraband excitations; furthermore, they gradually grow in the increment of ${n^{c(v)}}$. This
leads to an increase in the range of interband excitation frequency or
the peak width in D$_{J}(\omega ,B_{0})$. This means that the effective $k_{z}$-ranges gradually grow as the
frequency increases (Fig. 1(b)--(d)), and so do the ranges of the interband
excitation energies.

\begin{figure}[htb]
\centering\includegraphics[width=10cm]{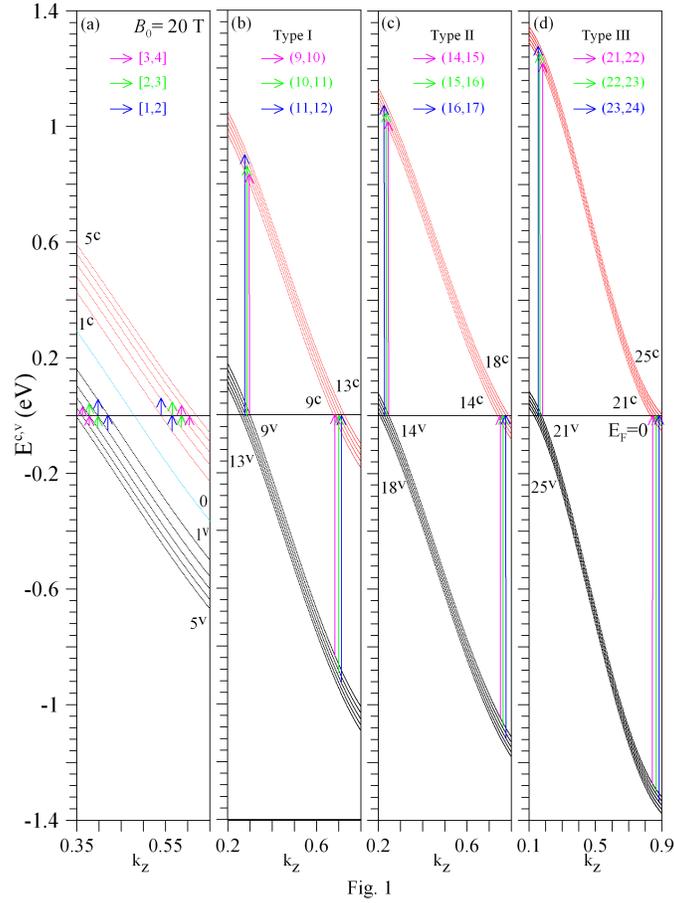}
\caption{The vertical optical excitations at $B_{0}=$ $20$ T due to the Landau subbands for (a) the intraband and (b)-(d) interband excitations. The allowed excitations are confined by two arrows of the same color. The square brackets and the parentheses represent the intraband and interband excitations, respectively.}
\end{figure}

 The joint density of states is the number of optical excitation channels,
which are directly reflected in the absorption spectra. The
spectral function can be expressed by the imaginary part of $\varepsilon (\omega
,B_{0})$, i.e., {\small $A(\omega )=$ }$\omega ^{2}${\small $\varepsilon
_{2}/2\pi $. Each D$_{J}(\omega
,B_{0})$ peak originates from excitations within a certain $k_{z}$-range
surrounded by the two arrows in Fig. 1. In each diagram, different colors are
used to denote excitation channels corresponding to their own D$_{J}$ peaks
 in Fig. 2. For the range $\omega <$ $0.1$ eV, the three
higher-frequency D$_{J}$ peaks of the intraband excitations, i.e., [1,2],
[2,3], and [3,4], are indicated by blue, green, and magenta curves,
respectively (Fig. 2(a)). These peaks partially overlap each other and
such an overlap might cause that the peaks merge together in the optical
spectrum, especially for the weak magnetic field cases. Moreover, each peak is a
composite structure of two intraband excitation channels: $[n^{c},(n+1)^{c}]$
and $[(n+1)$$^{v},n^{v}]$. The peak strength grows with an increasing
frequency because of the enlarged effective $k_{z}$-range.

\begin{figure}[htb]
\centering\includegraphics[width=10cm]{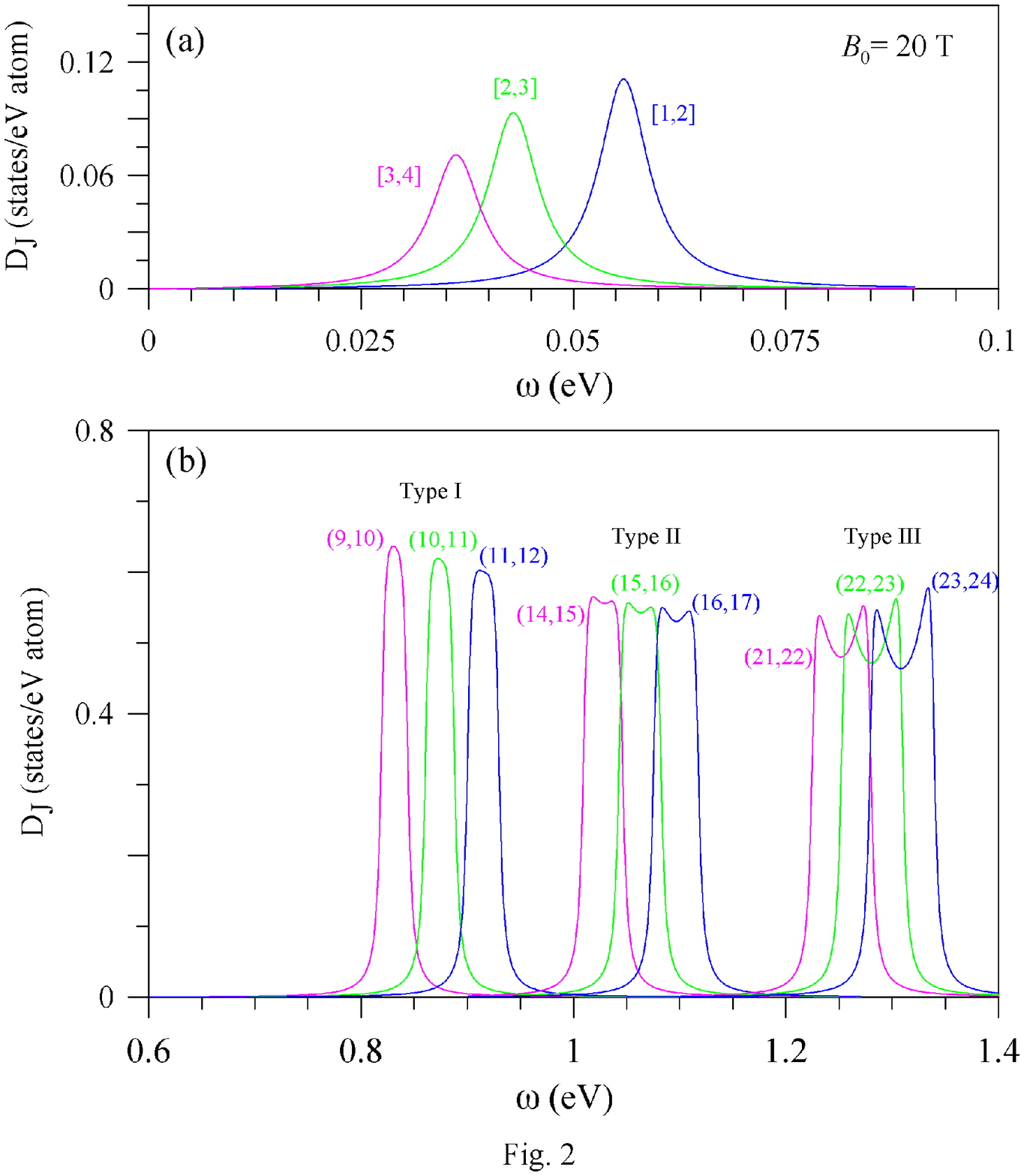}
\caption{The joint density of states corresponding to the Figure 1 for the (a) intraband (b) interband optical
excitations.}
\end{figure}

In the frequency range $\omega $ $\geq $ $0.1$ eV, the D$_{J}$
peaks are related to the interband excitations as shown in Fig. 2(b).
In the D$_{J}$, the three-peak structures associated with each
type exhibit different characteristics. This is a result of the fact
that the $k_{z}$-dependent curvature variations between two LSs in a
certain $k_{z}$ region are discrepant. The former three peaks
associated with the peaks of absorption
spectrum $A(\omega )$, are distinguishable. These Type I peaks are located within the range $0.1$
eV $<\omega <$ $1$ eV for a field strength of $B_{0}=$ $20$ T (indicated by the
red curve in Fig. 3(a)). Type II peaks are a combination of two neighboring
peaks, such as (14,15) and (15,16) or (15,16) and (16,17). Their composite peak intensity
is twice that of the original peaks. The peaks
correspond to the absorption spectrum with higher-intensity peaks in the
frequency range of $1$ eV $<\omega <$ $1.25$ eV (Fig. 3(a)). As to Type III, three neighboring peaks are merged to a single peak and its intensity is enhanced to almost
three times the original one. The D$_{J}$ peaks show up in the spectral frequency range
of $1.25$ eV $<\omega <$ $1.45$ eV. The higher-intensity absorption peaks
are caused by the merging of yet more D$_{J}$ peaks as a result of multiple channel
excitations. This clearly indicates that absorption peak height is increased
with the frequency following the sequence specified by the $\omega $-%
 range.

  In the absence of a magnetic field, the optical absorption
spectrum is indicated by the green curve in Fig. 3(a) and only a shoulder structure exists at $\omega $ $%
\approx $ $1.5$ eV \cite{CWChiu}. The magneto-optical spectrum demonstrates an abundance of
absorption peaks. It is dominated by the intraband and interband
excitations. The former lead to a stronger threshold peak and some weaker
peaks ($\omega <$ $0.1$ eV for $B_{0}=$ $20$ T), while the many groups with
a similar beating structure are attributed to the latter. The beating
oscillations are mainly determined by the joint density of states, since the
velocity matrix element is almost independent of the wave vector ($%
D_{m}\simeq 3$$\gamma _{0}$ $b^{\prime }/2$). The width of the absorption
peak gradually grows as the frequency increases. This leads to a higher
degree of overlap for the neighboring two peaks, thus the Group I
beating oscillations are formed within the range of $0.1$ eV $<\omega <$ $1$
eV. The frequency range associated with the other groups shifts to a higher
frequency following their group numbers. The oscillation is weaker towards the
end of each beating pattern. Moreover, these patterns are
reflected in the dielectric function. $A(\omega )$ exhibits a red shift in
the spectrum, more groups and weaker intensity as the field strength
decreases. The main reason is that the LS spacing, effective $k_{z}$-range
of the LS and the state degeneracy are lowered with a decreasing $B_{0}$.

\begin{figure}[htb]
\centering\includegraphics[width=11cm]{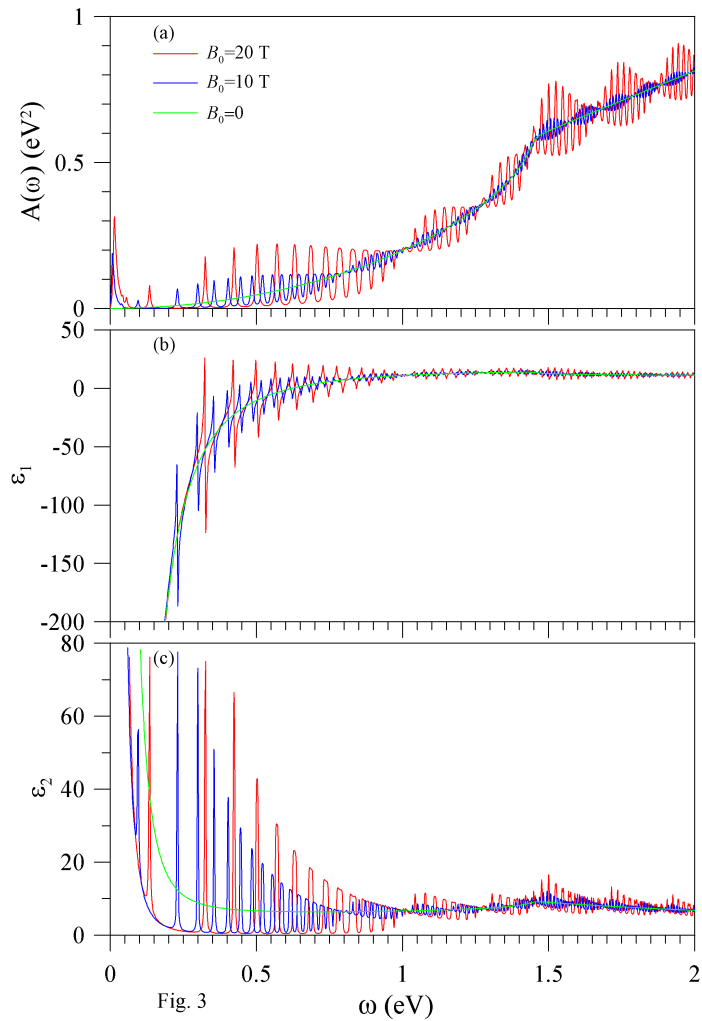}
\caption{(a) The optical absorption spectrum, (b) the real part, and
(c) the imaginary part of the dielectric function are shown for field
strengths 20 T, 10 T and zero.}
\end{figure}

 The single-particle excitations and collective plasmon modes can be
characterized by the real part $\varepsilon _{1}$ and the imaginary part $%
\varepsilon _{2}$ of the dielectric function. Each allowed LS excitation
produces a pair of asymmetric peaks in $%
\varepsilon _{1}$ and a peak in $\varepsilon _{2}$, as shown in Fig. 3(b) and (c), respectively. If the zero points in $%
\varepsilon _{1}$ occur where $\varepsilon _{2}$ vanishes, they are
associated with the undamped plasmon oscillations. The peak strength presents
a beating pattern in the dielectric function. Moreover, the magnitude
of $\varepsilon (\omega ,B_{0})$ should diminish at higher frequencies, owing
to the $\omega ^{-2}$ factor in Eq. (1). It should also be noted that $\varepsilon _{1}$
and $\varepsilon _{2}$, due to the intraband excitations, are quite large, e.g.,
their values are more than 10$^{3}$ times higher for $\omega <$ $0.05$ eV (not shown). Thus, they hardly
contribute to the loss fuction and reflectance spectra, and do not induce
prominent structures in such spectra. Moreover, the temperature only has an effect on the composite threshold peak
which is caused by the intraband excitations \cite{RBChen2}. As a result, the temperature effects are negligible in this work.

 The loss fuction, defined as ${\it {Im}}$[${-1/\epsilon (\omega ,B_{0})}$],
is useful for understanding the collective excitations that can be measured
by inelastic light and electron scattering spectroscopy \cite{CFChen, SMCollins,MFLin}.
 The loss fuction presents many noticeable peaks, as shown
in Fig. 4(a). These peaks are regarded as the collective excitations, only
coming from the interband excitations. The peak structures belonging to
Group I in terms of the energy range are prominent, whereas the other groups
belonging to different types own weak plasmon peaks. The higher intensity corresponds to a zero point
in $\varepsilon _{1}$ and a small value in $\varepsilon _{2}$ within the gap
region between two single-particle excitation energies, while the lower
intensity is subjected to strong Landau damping with a large $%
\varepsilon _{2}$. The plasmon peaks first rise, and then decline with
increasing LS transition channels. These peaks are gradually red-shifted and
diminished with respect to the decrease of field strength.

 The field-dependent plasmon frequency deserves a closer
investigation in order to understand the LS features. The frequency of each
plasmon structure grows with increasing field strength, as shown in Fig.
4(b). Plasmon peaks hardly survive in the loss function for a sufficiently
weak field strength. The low critical field occurs at higher LS transition
channels associated with a higher plasmon energy, while the lower LS
transition channels are subjected to the high critical field. It is
relatively easy to observe the plasmon peak for larger $B_{0}$ and $\omega $%
. Due to the state degeneracy and effective $k_{z}$-range of LS being
proportional to $B_{0}$, the low field strength only presents a few of
plasmon peaks from the higher LS transition channels. As for $B_{0}=0$, one
prominent plasmon peak arises at $\omega _{p}=0.63$ eV (Fig. 4(a)).

\begin{figure}[htb]
\centering\includegraphics[width=11cm]{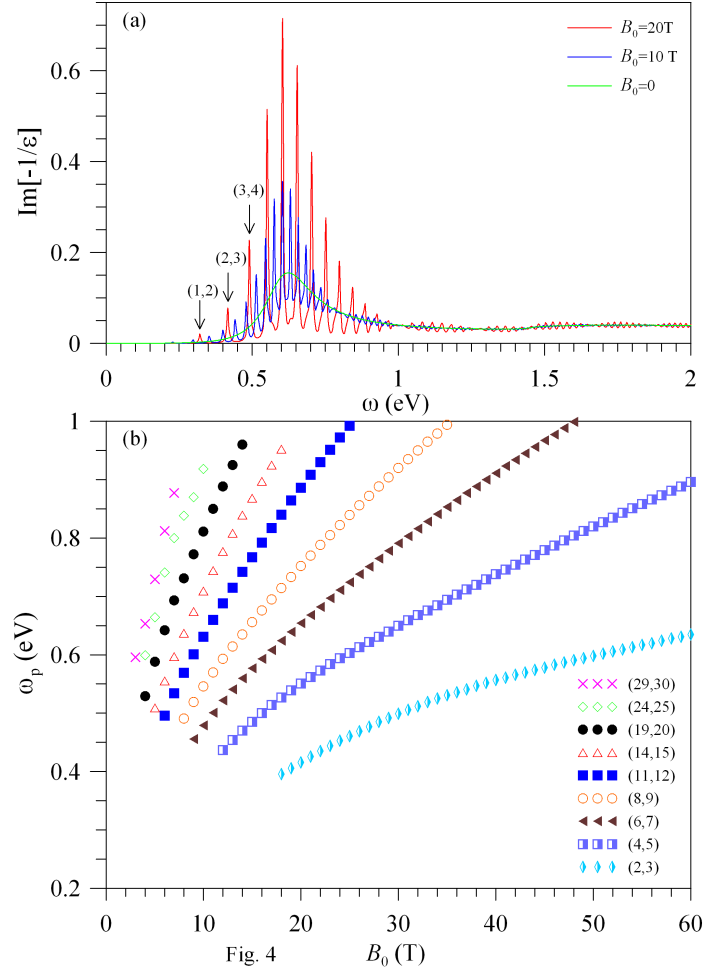}
\caption{(a) The loss function with respect to three different field
strengths. (b) Plasmon frequencies versus field strengths. Each curve
corresponds to a specific interband excitation.}
\end{figure}

 The magnetoreflectance calculated from the $R(\omega
,B_{0})=\left\vert 1-\sqrt{\epsilon (\omega ,B_{0})}\right\vert
^{2}/\left\vert 1+\sqrt{\epsilon (\omega ,B_{0})}\right\vert ^{2}$ clearly
depicts the features of single-particle and collective excitations. The
magnetoreflectance \ spectrum presents a series of field-dependent
oscillations in the beating pattern, compared with the featureless $%
R(\omega ,B_{0})$ at $B_{0}=0$ (Fig. 5(a)). The strongest beating pattern is located in the Group I. The magnetoreflectance spectrum contains both shoulders (peaks)
and dips, respectively, indicating the single-particle and collective excitations.
When the electromagnetic wave propagates in SHG, it is
attenuated very rapidly for strong single-particle excitations with a
very large $\varepsilon _{2}$, and most of the electromagnetic power is reflected.
On the other hand, if the electromagnetic wave frequency is the same as or higher than $\omega
_{p} $, the electromagnetic power is effectively absorbed by the collective
excitations, resulting in the plasmon dip structures of the spectrum.
The beating pattern diminishes its intensity and exhibits a
red shift as the field strength decreases.

\begin{figure}[htb]
\centering\includegraphics[width=10cm]{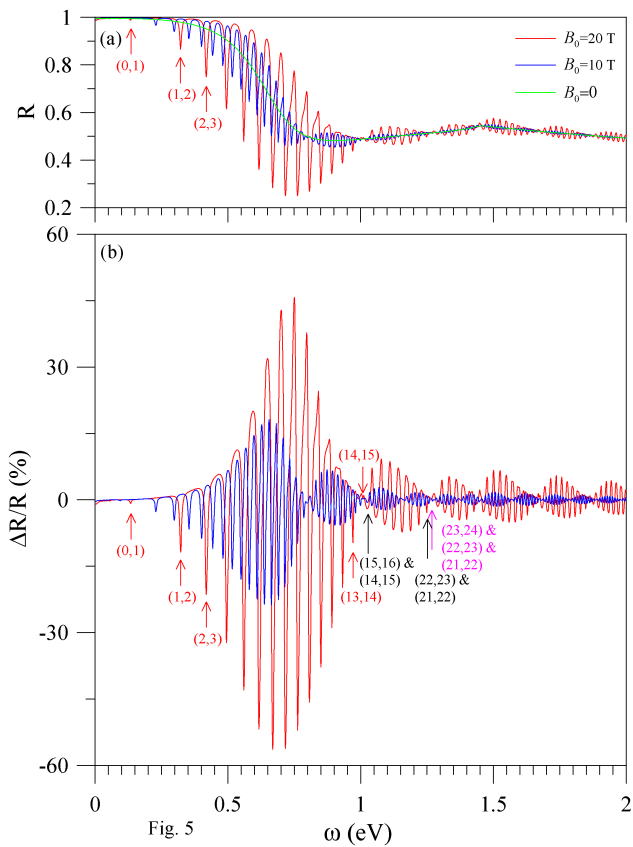}
\caption{(a) The magnetoreflectance spectrum and (b) the relative
magnetoreflectance spectrum for different field strengths.}
\end{figure}

 The relative magnetoreflectance, $\Delta R/R$ $=(R(\omega ,B_{0})-$
$R(\omega ,0))/R(\omega ,0)$ exhibits rather strong oscillations in the
beating pattern for different frequency ranges, as shown in Fig. 5(b). The
beating patterns of Group I present stronger intensities and
broader energy range than those of other groups. The strongest $\Delta R/R$ of the beating pattern of Group I is
close to $\pm 50\%$ for $B_{0}=20$ T, but below $\pm 10\%$ for the other
groups. These results could all be verified by experminental measurements. As the
field strength decreases, the energy range of each beat pattern is decreased. It
should be noted that up to now, the estimated variation ranges of $\Delta R/R$
 are all less than $20\%$ for other systems at the same field strength 20 T, e.g., Bernal graphite \cite{ZQLi, LCTung, WWToy} and YBa$_{2}$Cu$_{3}$O$_{y}$ \cite{WJPadilla}. SHG is predicted to exhibit the largest variations in its magnetoreflectance spectrum among all condensed-matter systems.

 The $\Delta R/R$ with respect to the field strength reveals
different groups of oscillation structures\ which could be tuned by the
photon energies (Fig. 6). The shoulders (peaks) and dips are attributed to
interband excitations. The oscillations still appear in the comparatively
weak field associated with the higher LS transition channels. These results
indicate that the oscillations
move more rapidly to higher $B_{0}$ as the photon energy increases. In the
range of $\omega <$ $0.5$ eV, all oscillations belong to Group I (Fig.
6(a)). The Group I and Group II oscillations coexist within the range of $%
0.5 $ eV $<\omega <$ $0.65$ eV (Fig. 6(b)). Thus, when photon energy
increases, $\Delta R/R$ exhibits more groups, as shown in Fig. 6(c).
Evidentally, there exists a relationship between the critical photon
energies and the group numbers, as these energies are identified to be $0.5$
eV, $0.65$ eV, $0.85$ eV, etc.. This phenomenon is a unique
characteristic of SHG and can be used to distinguish it from other
graphene-related systems.

 The oscillation structures of SHG are very different from
those of the other graphites. SHG presents a few groups of oscillation
structures with very strong $\Delta R/R$. The photon energy can be used to
modulate the total number of groups as a result of the strong $k_{z}$%
-dependent LS dispersion and the wide overlapping ranges of energy for different
transition channels. On the contrary, only low-energy irregular oscillation
structures was found in BG \cite{ZQLi, LCTung, WWToy, WJPadilla}.
 The low-lying LS dispersions with a narrow energy width of $\sim 0.2$ eV are responsible for the absence of beating oscillations. Up to now, no theoretical
calculations or experiments on the oscillation structures of RHG have been performed.
Based on the very weak LS dispersions and the lack of overlapping energy ranges of the
different transition channels, RHG supposedly
will only exhibit irregular oscillations of low intensities that will not demonstrate a beating pattern.

\begin{figure}[htb]
\centering\includegraphics[width=11cm]{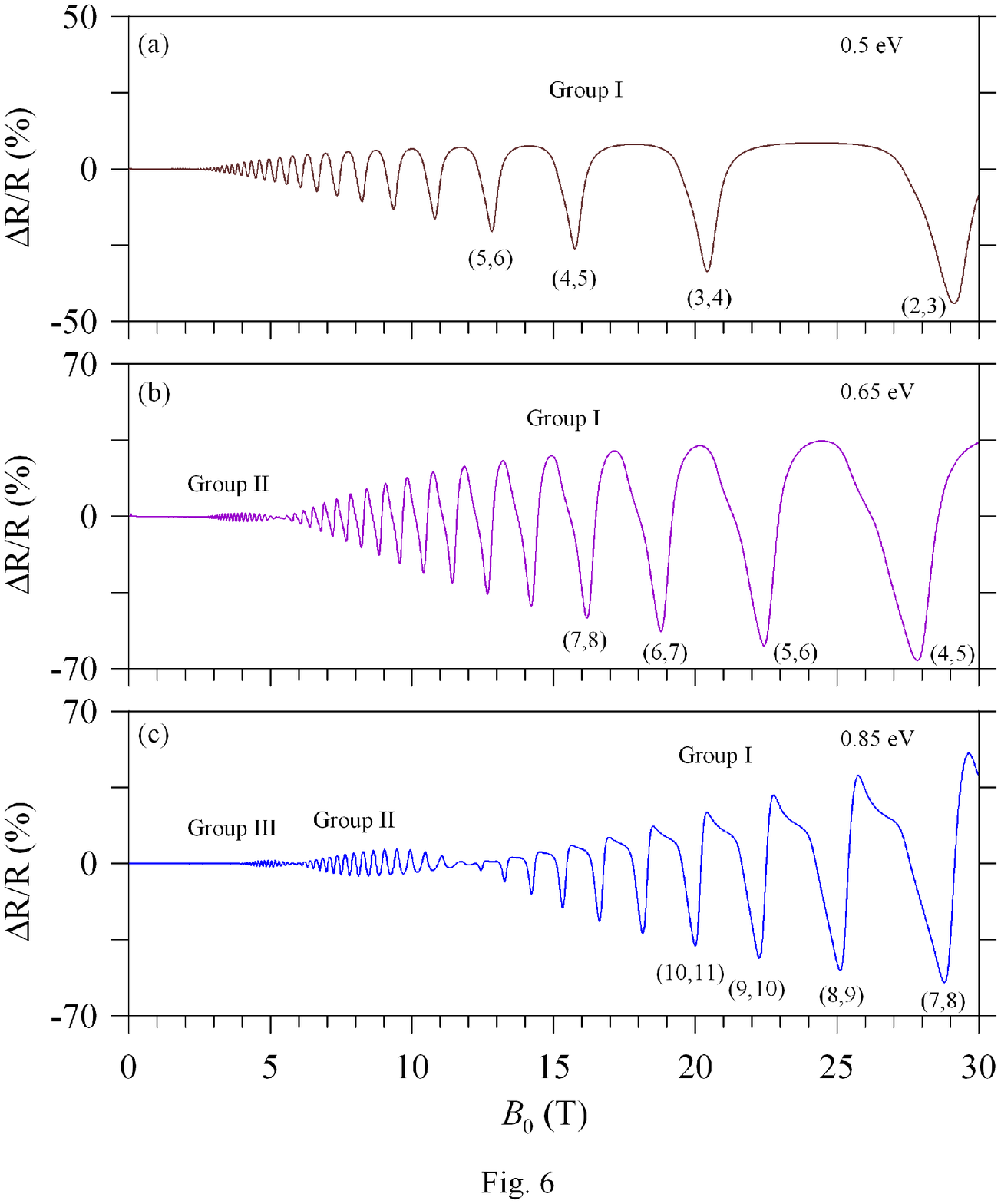}
\caption{The relative magnetoreflectance with respect to the field strength
reveals distinct groups of oscillation structures. Panels (a)-(c) represent
different photon energies.}
\end{figure}

\vskip 0.6 truecm
\par\noindent
{\bf 4. Conclusion }
\vskip 0.3 truecm


The magneto-optical properties of SHG demonstrate rich spectra with
beating structures, owing to the strong $k_{z}$-dependent LS and the wide
overlapping ranges of energy for different transition
channels. The single-particle excitations and collective excitations appear
simultaneously in the magneto-optical spectra and can be precisely
identified. The beating pattern of both the loss
function and reflectance in Group I exhibit stronger intensities and wider energy ranges than those in other groups. As field strength increases, the plasmon peaks of the loss function and the
dips of the reflectance are intensified, but the number of groups of beating structures is diminished. Moreover, the photon energy can modulate the
total number of groups of the oscillation structures. The unique magneto-optical properties, the beating oscillations and the largest variation in reflectance spectrum, could be confirmed by magneto-optical
spectroscopy measurements \cite{ZQLi, LCTung, EAHenriksen, ICrassee}.

\par\noindent {\bf Acknowledgments}

This work was supported by the NSC and National Center for Theoretical Sciences of Taiwan, under the
Grant No. NSC 98-2112-M-006-013-MY4.

\newpage
\renewcommand{\baselinestretch}{0.2}

\newpage \centerline {\Large \textbf {FIGURE CAPTIONS}}

\vskip0.5 truecm 

Fig. 1. The vertical optical excitations at $B_{0}=$ $20$ T due to the Landau subbands for (a) the intraband and (b)-(d) interband excitations. The allowed excitations are confined by two arrows of the same color. The square brackets and the parentheses
represent the intraband and interband excitations, respectively.

 Fig. 2. The joint density of states corresponding to the Fig. 1 for the (a) intraband (b) interband optical
excitations.

 Fig. 3. (a) The optical absorption spectrum, (b) the real part, and
(c) the imaginary part of the dielectric function are shown for field
strengths 20 T, 10 T and zero.

 Fig. 4. (a) The loss function with respect to three different field
strengths. (b) Plasmon frequencies versus field strengths. Each curve
corresponds to a specific interband excitation.

 Fig. 5. (a) The magnetoreflectance spectrum and (b) the relative
magnetoreflectance spectrum for different field strengths.

 Fig. 6. The relative magnetoreflectance with respect to the field strength
reveals distinct groups of oscillation structures. Panels (a)-(c) represent
different photon energies.

\end{document}